\documentclass[10pt,final,journal,a4paper,oneside,twocolumn]{IEEEtran}

\usepackage{float}
\usepackage{srcltx}
\usepackage{psfrag}
\usepackage{epsfig}
\usepackage{color}
\usepackage[tbtags,sumlimits,nointlimits,reqno]{amsmath}
\usepackage{amssymb}
\usepackage{showkeys}   
\usepackage{color}
\usepackage{float}
\usepackage{mathtools, cuted}
\usepackage[makeroom]{cancel}
\usepackage{relsize}
\usepackage{bbm}
\usepackage{ulem}
\usepackage[dvipsnames]{xcolor}
\usepackage{cite}

\makeatletter
\let\MYcaption\@makecaption
\makeatother

\usepackage[font=footnotesize]{subcaption}

\makeatletter
\let\@makecaption\MYcaption
\makeatother

\renewcommand{\r}{\mathbf{r}}
\newcommand{\rhob}{\boldsymbol{\rho}}
\newcommand{\E}{\mathbf{E}}
\newcommand{\J}{\mathbf{J}}
\renewcommand{\H}{\mathbf{H}}
\newcommand{\K}{\mathbf{K}}
\newcommand{\X}{\mathbf{X}}
\newcommand{\M}{\mathbf{M}}
\renewcommand{\L}{\mathbf{\mathcal{L}}}
\newcommand{\R}{\mathbf{\mathcal{R}}}

\newcommand{\nh}{\mathbf{\hat n}}

\newcommand{\0}{\varnothing}
\newcommand{\I}{\mathbb{I}}
\newcommand{\Ev}{\mathbb{E}}
\newcommand{\Jv}{\mathbb{J}}
\newcommand{\Hv}{\mathbb{H}}
\newcommand{\Kv}{\mathbb{K}}
\newcommand{\SLm}{\mathbb{S}^\L}
\newcommand{\SRm}{\mathbb{S}^\R}
\newcommand{\SLmOne}{\mathbb{S}^{\L1}}
\newcommand{\SRmOne}{\mathbb{S}^{\R1}}
\newcommand{\SLmTwo}{\mathbb{S}^{\L2}}
\newcommand{\SRmTwo}{\mathbb{S}^{\R2}}
\newcommand{\Sm}{\mathbb{S}}
\newcommand{\Ym}{\mathbb{Y}}
\newcommand{\Bm}{\mathbb{B}}

\newcommand{\Chibb}{\mathbb{X}}
\newcommand{\N}{\mathbb{\hat N}}

\newcommand{\tjs}[1]{\textcolor{Black}{#1}}

\newcommand{\tjsrm}[1]{\textcolor{Black}{}}
\newcommand{\tjsrp}[2]{\textcolor{Black}{}{#2}}
\newcommand{\sg}[1]{\textcolor{Black}{#1}}

\begin{document}
\normalem

\title{Scattering Field Solutions of Metasurfaces based on the Boundary Element Method (BEM) for Interconnected Regions}

\author{%
       Scott A. Stewart, Sanam Moslemi-Tabrizi, Tom. J. Smy and Shulabh~Gupta
\thanks{S. Stewart, S. Moslemi-Tabrizi, T. J. Smy and S.~Gupta, are with the Department of Electronics, Carleton University, Ottawa, Ontario, Canada. Email: ShulabhGupta@cunet.carleton.ca}
}
\markboth{MANUSCRIPT DRAFT}
{Shell \MakeLowercase{\textit{et al.}}: Bare Demo of IEEEtran.cls for Journals}

\maketitle
\begin{abstract}
\tjs{A methodology for determining the scattered Electromagnetic (EM) fields present for interconnected regions with common metasurface boundaries is presented. The method uses a Boundary Element Method (BEM) formulation of the frequency domain version of Maxwell's equations -- which expresses the fields present in a region due to surface currents on the boundaries. Metasurface boundaries are represented in terms of surface susceptibilities which when integrated with the Generalized Sheet Transition Conditions (GSTCs), gave rise to an equivalent configuration in terms of electric and magnetic currents. Such a representation is then naturally incorporated into the BEM methodology. Two examples are presented for EM scattering of a Gaussian beam to illustrate the proposed method. In the first example, metasurface is excited with a diverging Gaussian beam, and the scattered fields are validated using a semi-analytical method. Second example concerned with a non-uniform metasurface modeling a diffraction grating, whose results were confirmed with conventional Finite Difference Frequency Domain (FDFD) method.}

\end{abstract}



\section{Introduction}

Metasurfaces are 2D counterparts of more general volumetric metamaterials\cite{meta2}. They are composed of 2D arrays of sub-wavelengths unit cells whose microscopic response can be tailored to engineer the macroscopic response of the metasurfaces. By controlling these susceptibilities, various sophisticated wave transformations can be achieved. With recent intense activity in this area wide variety of metasurfaces have been developed enabling versatile wave transforming applications ranging from Radio Frequencies (RF) to optics \cite{MetaHolo, MetaCloak, MetaFieldTransformation, ReconfgMSoptics, ShaltoutSTMetasurface}.

An important problem in metasurface research is the development of fast, efficient and reliable simulation platforms. While the metasurfaces are composed of sub-wavelength resonating cells, they themselves are typically large compared to the wavelengths of operation, so that their numerical simulation is essentially a multi-scale problem. To address this issue, the representation of a physical metasurface is transformed to an ideal zero-thickness model, which is expressed using tensorial effective surface susceptibility densities, $\bar{\bar{\chi}}$ to account for various Electromagnetic (EM) effects, including bi-anisotropy \cite{Metasurface_Synthesis_Caloz}. An equivalent zero thickness model of the metasurface represents a spatial discontinuity and thus is treated using Generalized Sheet Transition Conditions (GSTCs) \cite{KuesterGSTC,IdemenDiscont, GSTC_Holloway}. Based on GSTCs and the surface susceptibilities, various numerical techniques have been proposed recently to solve for the scattered fields from the metasurface, for both frequency \cite{Caloz_MS_Siijm, CalozFDTD} and time-domain \cite{Vaheb_FDTD_GSTC, Smy_Metasurface_Space_Time, FDTD_Metasurface_Disperive} simulations using Finite Difference (FD) methods, Finite Element Methods (FEM) \cite{Caloz_FEM} and Integral Equations (IE) in the Spectral Domain (SD) \cite{Caloz_Spectral}.

All these numerical methods have been demonstrated for computing the scattered fields from standalone metasurfaces, and are practically suitable for finite regions of spaces only. For cases where the metasurface is placed with various other scatterers as part of an electrically large system these methods become computationally challenging requiring substantial memory and computational resources. To solve such problems more generally, Boundary Element Methods (BEM) have been developed (producing a vast body of literature) which solve the scattered EM fields in terms of physical and equivalent electric and magnetic surface currents in a given volume of interest using integral form of the Maxwell's equations \cite{chew2009integral,ComputationalBEM, AppBEMEM, OpticalBEM, FE_BEM_Impedance}. As the BEM method does not require meshing of the entire volume, it is computationally efficient and thus well suited for solving electrically large problems. 

In this work, an idealized model of the zero thickness metasurface in conjunction with the GSTCs is treated as a generalized boundary condition connecting different volumetric regions, and integrated into the BEM to solve for total scattered fields in electrically large computational domains. Given the generalized field transformation properties of the metasurface, there are non-zero electric and magnetic surface currents that exist on the metasurface. Compared to conventional BEM methods based on Electric Field Integral Equations (EFIE) and Magnetic Field Integral Equations (MFIE) which typically involve only electric surface currents the proposed method thus solves for the total fields in the presence of both electric and magnetic equivalent currents on the metasurface.

The paper is structured as follows. Sec. II outlines the general problem consisting of arbitrary number of finite space regions connected through various conventional boundaries in addition to a set of metasurface boundaries. Sec. III presents the BEM procedure for discretizing Maxwell's equations. Sec. IV integrates the metasurface boundaries into the BEM, and provides an illustrative example of two regions connected by a single metasurface. Various numerical results are presented in Sec. V followed by conclusions in Sec. VI.

\section{Problem Formulation}

\subsection{General Scattering Problem \& Conventional Boundary Conditions}

\begin{figure}
    \centering
    \centering
\psfrag{a}[c][c][0.8]{$\mathbf{E}_{s,1},\; \mathbf{H}_{s,1}$}
\psfrag{b}[c][c][0.8]{$\mathbf{E}_{s,2},\; \mathbf{H}_{s,2}$}
\psfrag{c}[c][c][0.8]{$\mathbf{E}_{s,3},\; \mathbf{H}_{s,3}$}
\psfrag{d}[c][c][0.8]{$\mathbf{E}_{s,4},\; \mathbf{H}_{s,4}$}
\psfrag{e}[c][c][0.8]{$\mathbf{E}_{s,n},\; \mathbf{H}_{s,n}$}
\psfrag{f}[c][c][0.8]{$\infty$}
\psfrag{g}[c][c][0.8]{$\J_0,~\K_0$}
\psfrag{h}[c][c][0.8]{$\epsilon,\;\mu$}
\psfrag{k}[c][c][0.6]{$\mathbf{J_s}=0$}
\psfrag{m}[c][c][0.6]{$\mathbf{K_s}=0$}
\includegraphics[width=0.95\columnwidth]{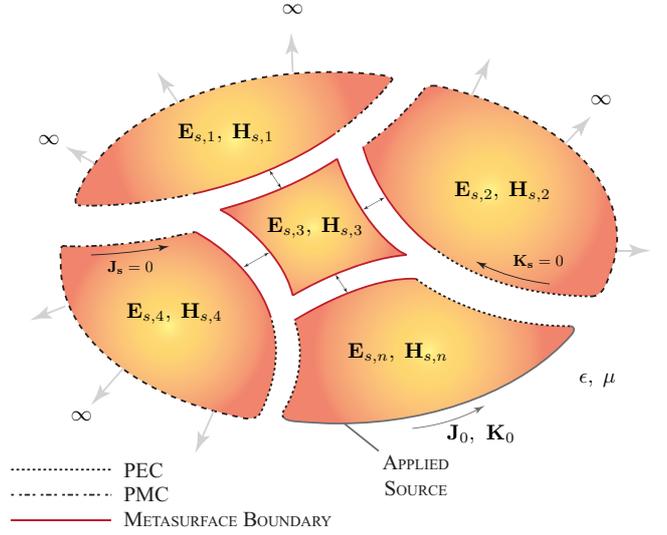}

%
\caption{Generalized field scattering problem. Illustration showing an exploded view of a field scattering problem consisting of arbitrary number of finite space regions connected through various boundaries such as Perfect Electric Conductors (PECs) and Perfect Magnetic Conductors (PMCs), in addition to a set of different metasurface boundaries. A known source is also applied at one or some of the boundaries. \label{fig:FS_Problem}
}
\end{figure}

Consider a general field scattering problem illustrated in Fig.~\ref{fig:FS_Problem}, consisting of several volumetric regions of space. Each of the regions of interest are connected to their neighboring regions through various EM boundary conditions, which could represent either physical or purely mathematical boundaries in space. A known source is next applied at one (or more) boundaries in terms of electric and magnetic surface currents, $\mathbf{J_0},~\mathbf{K_0}$ for instance, which produces scattered fields throughout all the regions. The objective here is to compute the total scattered fields in various regions, satisfying all the boundary conditions and solving the Maxwell's equations self-consistently.

Two boundaries commonly encountered are Perfect Electric Conductors (PECs) and Perfect Magnetic Conductors (PMCs), which are impenetrable boundaries where the total EM fields goes to zero. For instance, in case of a PEC boundary, the tangential E-fields are continuous and zero, while the H-fields are discontinuous giving rise to surface electric currents $\J_s$. Similarly for a PMC boundary, the tangential H-fields are continuous and zero, while the E-fields are discontinuous giving rise to surface magnetic currents $\K_s$. Formally, 
\begin{subequations}
\begin{equation}
-\nh \times  \H_{s,1} = \J_s,
\end{equation}
\begin{equation}
\nh \times  \E_{s,1} = \K_s.
\end{equation}
\end{subequations}
\noindent for PEC and PMC boundaries, respectively, where $\H_{s,1}$ and $\E_{s,1}$ are the total H- and E-fields on the boundary in the incident region of the boundaries and $\nh_i $ is the unit normal vector to the surface at given point.

For other general boundaries for which the tangential E- and H-fields are both discontinuous (i.e. penetrable boundaries), both electric and magnetic surface currents exist and are given by

\begin{subequations}\label{Eq:J_K}
\begin{equation}
	-\nh \times \left(\E_{s,1} - \E_{s,2}\right) = \K_s
	\end{equation}
	\begin{equation}
	\nh \times \left(\H_{s,2} - \H_{s,1}\right) = \J_s,
	\end{equation}
\end{subequations}

\noindent where a special case is the interface between two dielectric materials where both $\K_s$ and $\J_s$ are zero.  Completely general boundaries with non-zero $\K_s$ and $\J_s$ can be realized using \emph{Electromagnetic Metasurfaces}  and described in terms of effective surface polarization densities as described next. 

\subsection{Metasurface Boundaries - Generalized Sheet Transition Conditions (GSTCs)}
\begin{figure}
\centering
\psfrag{a}[c][c][0.8]{$\bar{\bar{\chi}}_\text{ee}$,~$\bar{\bar{\chi}}_\text{mm}$}
\psfrag{c}[c][c][0.8]{$\mathbf{J}_s,~\mathbf{K}_s$}
\psfrag{b}[r][c][0.8]{$\mathbf{E}_{s,1},~\mathbf{H}_{s,1}$}
\psfrag{j}[c][c][0.8]{$\mathbf{E}_{s,2},~\mathbf{H}_{s,2}$}
\psfrag{k}[c][c][0.8]{$\epsilon,~\mu$}
\psfrag{d}[c][c][0.8]{$\nh$}
\psfrag{e}[c][c][0.8]{$\delta =0$}
\includegraphics[width=0.7\columnwidth]{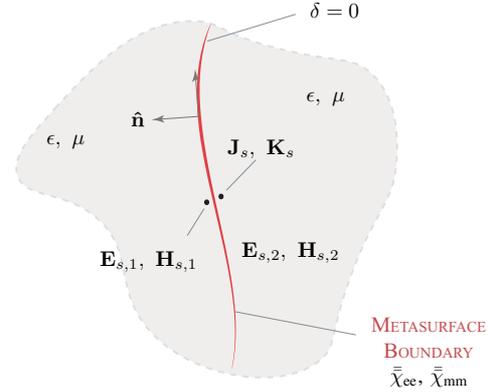}

\caption{Metasurface as a general boundary, described in terms of its surface susceptibilities $\bar{\bar{\chi}}$s.\label{fig:MS_BC}}
\end{figure}
A metasurface is a two dimensional array of sub-wavelength electromagnetic scatterers with zero thickness ($\delta=0$) which produces a spatial discontinuity in the amplitude and phase of an incoming electromagnetic wave, as shown in Fig.~\ref{fig:MS_BC}. The Generalized Sheet Transition Conditions (GSTCs) were developed by Idemen in  \cite{IdemenDiscont} to model such these discontinuities and were later applied to metasurfaces in \cite{KuesterGSTC}. For a general metasurface embedded inside a uniform media with $(\epsilon, \mu)$, the GSTCs can be written in the frequency-domain as:
\begin{subequations} \label{Eq:FullGSTC}
\begin{equation}
	 \nh \times \Delta \H  = j\omega \mathbf{P}_{||} - \nh \times \nabla_{||}\M_{n}
\end{equation}
\begin{equation}
	\nh \times \Delta \E = -j\omega\mu_0 \M_{||} - \nh \times \nabla_{||}\left(\frac{\mathbf{P}_{n}}{\epsilon_0}\right)
\end{equation}	
\end{subequations}
\noindent where $\Delta\psi=\psi_2 - \psi_1$ represents the difference between the fields across the metasurface, and $\mathbf{P}$ and $\M$ are the electric and magnetic surface polarization densities. The term $X_{||}$ is the component that is tangential to the metasurface and the term $X_{n}$ is perpendicular to the metasurface. The surface polarization densities are produced in response to a field interacting with the metasurface. These polarizations can be related to the average electric and magnetic fields through the use of surface susceptibilities, and are expressed in general as:
\begin{subequations}
\begin{equation}
	 \mathbf{P}(\omega) = \epsilon \overline{\overline{\chi}}_\text{ee} \E_\text{avg}(\omega) + \sqrt{\mu\epsilon}\overline{\overline{\chi}}_\text{em}\H_\text{avg}(\omega)
\end{equation}
\begin{equation}
	\M(\omega) = \sqrt{\frac{\epsilon}{\mu}}\overline{\overline{\chi}}_\text{me}\E_\text{avg}(\omega) + \overline{\overline{\chi}}_\text{mm} \H_\text{avg}(\omega)
\end{equation}	
\end{subequations}
\noindent where $\E_\text{avg}=(\E_{s,2}+\E_{s,1})/2$ and $\H_\text{avg}=(\H_{s,2}+\H_{s,1})/2$ are the average tangential electric and magnetic field across the metasurface respectively, expressed in terms of the total fields in each of the two regions across the boundaries. $\overline{\overline{\chi}}_\text{ee}$ and $\overline{\overline{\chi}}_\text{mm}$ are the effective electric and magnetic surface susceptibilities respectively, and $\overline{\overline{\chi}}_\text{em}$ and $\overline{\overline{\chi}}_\text{me}$ are the cross-anisotropic surface susceptibilities of the metasurface. Let us assume for simplicity that $\mathbf{P}_{n}=\M_{n}=0$ which simplifies Eq.~\eqref{Eq:FullGSTC} and yields:
\begin{subequations}\label{Eq:ExpandedGSTC}
\begin{equation}
	 \nh \times \Delta \H  = j\omega\epsilon\overline{\overline{\chi}}_\text{ee} \E_\text{avg} + j\omega\sqrt{\mu\epsilon}\overline{\overline{\chi}}_\text{em}\H_\text{avg}\label{Eq:ExpandedGSTC1}
\end{equation}
\begin{equation}
	\nh \times \Delta \E = -j\omega\sqrt{\mu\epsilon}\overline{\overline{\chi}}_\text{me}\E_\text{avg} -j\omega\mu\overline{\overline{\chi}}_\text{mm} \H_\text{avg}.\label{Eq:ExpandedGSTC2}
\end{equation}	
\end{subequations}

The surface susceptibilities thus set the relationship between all the scattered fields across the metasurface, which can alternatively be synthesized to transform specified incident fields into desired transmission and reflection fields, i.e. total scattered fields \cite{MS_Synthesis}. Therefore, the surface susceptibility description of metasurfaces represents a powerful platform to describe arbitrary boundary conditions. 

\subsection{Scattering Formulation}

All the boundary conditions above relate the surface currents to the fields just across the boundaries.
However, the general goal of the scattering problem illustrated in Fig.~\ref{fig:FS_Problem} is determining the total scattered fields anywhere in the entire computational region.
\tjs{The approach taken here is to determine the fields within each region using an integral representation of Maxwell's equations in the frequency domain. Within each region the fields are a consequence of the surface currents ($\K_s$ and $\J_s$) present on that regions boundaries. For impenetrable boundaries the surface currents are determined by a boundary condition such as a PEC or PMC. For penetrable boundaries ({\em interfaces}) the currents are such that the interface conditions are maintained. As interfaces allow coupling between regions, the surface currents on these boundaries contribute to the fields in both regions. Some regions will have external boundaries that extend to infinity and allow for free radiation. These boundaries are handled naturally by the integral representation of the electromagnetic equations due to the use of Green's function that goes to zero at infinity.}

\tjs{Below we will show how, using the BEM method, the integral equations in each region are coupled through the interfaces to form a complete set of self-consistent linear equations that can be solved for the surface currents present for all regions. It should be noted that the \sg{resulting} surface currents are not typically physical but mathematical artifacts that enforce the boundary conditions and capture the geometrical implications of the field configuration for each region.} 


\section{Boundary Element Method (BEM)}

\subsection{Integral Equations for the Regions}

When applied to EM the Boundary Element Method (BEM) uses an integral representation of Maxwell's equations to determine the scattered fields inside of a region. It is assumed that electromagnetic fields are produced by electric and magnetic surface current densities, $\mathbf{J}$ and $\mathbf{K}$, present on the surfaces enclosing a uniform volume of space. These surface current densities are integrated over the entire surface using the frequency domain version of Maxwell's equations, giving \cite{Method_Moments,MoM_Thesis}:
\begin{subequations}
\begin{align}
	\E(\r) = &-j\omega\mu\iint_{S}G(\r,\r')\left[1+\frac{1}{k^2}\nabla'\nabla'\cdotp\right]\J^s(\r') \,d\r' \label{Eq:ScatteredEField}\\
			&- \nabla \times \iint_{S}G(\r,\r')\K^s(\r') \,d\r'\nonumber\\
	\H(\r) = &-j\omega\epsilon\iint_{S}G(\r,\r')\left[1+\frac{1}{k^2}\nabla'\nabla'\cdotp\right]\K^s(\r') \,d\r' \label{Eq:ScatteredHField}\\
			&+ \nabla \times \iint_{S}G(\r,\r')\J^s(\r') \,dr'\nonumber
\end{align}
\end{subequations}
\noindent where $\E(\r)$ and $\H(\r)$ are the electric and magnetic fields inside of the region enclosed by the integration, $k$ is the wave vector in the region, $G$ is the Green's function for electrodynamics and $\J^s$ and $\K^s$ are the electric and magnetic surface current densities on the surface, and $\r = (x,y,z)$. Time convention used here is $e^{j\omega t}$. Primed and unprimed variables refer to source and observation locations, respectively. Because Eq.~\eqref{Eq:ScatteredEField} and \eqref{Eq:ScatteredHField} have similar terms, it is useful to represent these equations using a sum of linear operators acting on the surface currents as
\begin{subequations} \label{Eq:SimpleScatteredField}\
\begin{align}
	\E(\r) &= -j\omega\mu(\L\J^s)(\r) - (\R\K^s)(\r) \label{Eq:SimpleScatteredEField}\\
	\H(\r) &= -j\omega\epsilon(\L\K^s)(\r) + (\R\J^s)(\r), \label{Eq:SimpleScatteredHField}
\end{align}
\end{subequations}
\noindent where the operators $\L$ and $\R$ are written as:
\begin{subequations}\label{Eq:Operator}
\begin{align}
	(\L\X)(\r) &= \iint_{S}G(\r,\r')\left[1+\frac{1}{k^2}\nabla'\nabla'\cdotp\right]\X(\r') \,d\r'\label{Eq:LeftOperator}\\
	(\R\X)(\r) &= \nabla \times \iint_{S}G(\r,\r')\X(\r') \,d\r'. \label{Eq:RightOperator}
\end{align}
\end{subequations}
\noindent These operators involve the use of a Green's function which characterizes the impulse response of an inhomogeneous linear differential equation \cite{Advanced_Engineering_EM}. In the case of electrodynamics, the Green's function is a solution to the Helmholtz equation: 
\begin{align}
	\nabla^2G(\r, \r') + k^2G(\r, \r') = -\delta(\r, \r') \label{Eq:Helmholtz}
\end{align}
\noindent where $\delta$ is the delta function in space. The use of the Green's function in Eq. (\ref{Eq:SimpleScatteredField}) and (\ref{Eq:Operator}) represents the generation of the fields from the surface currents present at the interfaces. The linearity of the Maxwell's equations is next used to form the total field response from a superposition of various impulse responses where the source of each impulse is prescribed by the surface current distributions. The solution to Eq.~\eqref{Eq:Helmholtz} is well known and in two and three dimensions can be written as \cite{Method_Moments}:
\begin{subequations} 
\begin{align}
	G(\r,\r') &= \left(\frac{e^{-jk|\r-\r'|}}{4\pi|\r-\r'|}\right).\\
	G(\rhob, \rhob') &= -\frac{j}{4}H_0^{(2)}(k|\rhob - \rhob'|)\label{Eq:2DGreen}
\end{align}
\end{subequations}
%
%
\noindent where $\rhob = (x,y)$ is the position vector in two dimensions, and $H_0^{(2)}$ is the Hankel function of the first and second kind. \tjs{Note that the natural boundary condition for both of these functions is that the field is zero at infinity.}


\tjs{For situations with multiple connected regions as shown in Fig. \ref{fig:FS_Problem}, the integral equation operators (Eq. \ref{Eq:Operator}) can be applied to each region and interface equations are used to couple the regions. Each region is ``extracted'' and the \sg{scattered fields due to the corresponding surface currents are next computed by operators} $\L^i$ and $\R^i$ (for the i'th regions). \sg{These} surface currents present on the interfaces will be unknowns determined by the complete self-consistent solution of the entire domain.} 

\tjs{\sg{It should be noted} that within each domain the fields are created by currents present on the boundaries of that region \sg{only}. These currents represent the fields coupled in from the adjacent regions as determined by the interface equations or the imposition of a boundary condition. As such they are fictitious currents that: 1) enforce the interface/boundary conditions; 2) capture the geometry of the region; and 3) \sg{take into account} the influence of the fields of the surrounding region.}

\subsection{Discretization -- BEM with Pulse Functions}

The BEM is a well known and thoroughly researched method that calculates the scattered fields interacting with surfaces prescribing a region by descritizing Eq. (\ref{Eq:SimpleScatteredField}) \cite{Method_Moments,MoM_Thesis}. To model a region we assume that the surfaces of the region are discritized into a collection of elements -- where for our 2D case we would describe the surfaces/interfaces by line segments. In this paper it is assumed that there are two induced electric and magnetic surface currents ($\J_s$ and $\K_s$) present in the element; both of  which ``flow'' parallel to the surface and are pulse functions and uniform over the entire element\cite{Method_Moments}.\footnote{There are implementations of the BEM that use higher order interpolation for the surface currents over the boundary to improve the accuracy of the method \cite{MoM_Thesis,chew2009integral}, however, for simplicity we use \tjsrp{linear elements.}{uniform elements.}}

The surfaces prescribing the region are described by a set of $N$ line segments each centered at $\r_j$, with a surface normal $\nh_j$ and a length $d\r_j$ for the $j$'th segment. Under these assumptions Eq. (\ref{Eq:Operator}a) and (\ref{Eq:Operator}b) are descritized by evaluating the Green's function at the center of the line segment ($\r_j$) and using the length $d\r_j$ as the weight of the contribution to the sum giving:
\begin{subequations}\label{Eq:BEMOper}
\begin{align}
	(\L\X)(\r_i) &= \sum_{j=1}^N G(\r_i,\r_j)\left[1+\frac{1}{k^2}\nabla'\nabla'\cdotp\right]\X(\r_j) \,d\r_j  \nonumber\\
	&= \sum_{j=1}^N S^\L_{i,j} \X(\r_j)\\
	(\R\X)(\r_i) &= \nabla \times  \sum_{j=1}^N G(\r_i,\r_j)X(r_j) \,d\r_j  \nonumber\\
	& = \sum_{j=1}^N S^\R_{i,j} \X(\r_j), 
\end{align}
\end{subequations}
with 
\begin{subequations} \label{Eq:SCoeff}
\begin{align}
S^\L_{i,j} &= G(\r_i,\r_j)\left[1+\frac{1}{k^2}\nabla'\nabla'\cdotp\right] \,d\r_j\\
S^\R_{i,j} &= \nabla \times G(\r_i,\r_j) \,d\r_j.
\end{align}
\end{subequations}
\noindent These equations can then be used with Eq.~(\ref{Eq:SimpleScatteredField}) to relate the set of surface fields present at the elements ($\Ev^s$ and $\Hv^s$) to the set of surface currents ($\Jv^s$ and $\Kv^s$) by two matrix equations, 
\begin{subequations}\label{Eq:BEMScatteredField}
\begin{equation}
	\Ev^s = -j\omega\mu \SLm \Jv^s - \SRm \Kv^s
\end{equation}
\begin{equation}
	\Hv^s = -j\omega\epsilon \SLm \Kv^s + \SRm \Jv^s
	\end{equation}
\end{subequations}
\noindent where $\SLm$ and $\SRm$ are matrices that are formed using the $S^\L_{i,j}$ and $S^\R_{i,j}$ coefficients defined in Eq.~\eqref{Eq:SCoeff}. \tjs{It is evident that as we are determining the surface fields from the surface currents there is a contribution to each field from the self-same element. As the Green's function has a singularity at the source location when $\r = \r'$ this contribution needs to be handled carefully. Standard procedures exist for extracting this singularity and evaluating this contribution to the total field solution\cite{Method_Moments}.}

These equations thus relate the fields created to a known set of surface currents, however, in general (except for a defined source) the currents at the interfaces are unknown and \tjs{constrained} by various interface boundary conditions (BC) discussed in Sec.~II. The relationship between the scattered fields and the surface currents represented by Eq. (\ref{Eq:BEMScatteredField}) is \tjs{an additional requirement on the solution that the fields are a solution to Maxwell's equations}. The BEM method thus combines the various BC equations at every interface with Eq.~\eqref{Eq:BEMScatteredField} to solve for the unknown surface current distributions, from which the general scattered fields can finally be calculated. 

To illustrate this method, let us consider an arbitrarily shaped boundary described in terms of the electric and magnetic surface currents. These boundary conditions can be expressed in matrix form, once the interface is discretized so that
\begin{subequations}\label{Eq:GenBC}
\begin{align}
    -\N \left(\Ev_{s,2} - \Ev_{s,1}\right) &= \Kv_s\\
	\N \left(\Hv_{s,2} - \Hv_{s,2}\right) &= \Jv_s,
\end{align}
\end{subequations}
\noindent where $\N$ is a matrix operator formed from the operator $(\nh_i \times)$. In addition to these BC's an analysis will need to define a source. This can be done by defining a sub-set of the interface elements to be a source and prescribing a known electric and magnetic field distribution (or alternatively the electric and magnetic surface currents) on this portion of the surface, 
\begin{align}
   	\Ev_\text{so}  = \Ev_0,~ \Hv_\text{so}  = \Hv_0\label{Eq:Sou}
\end{align}

\begin{figure}
\begin{subfigure}[b]{0.23\textwidth}
\centering
\psfrag{a}[c][c][0.8]{$\hat{n}$}
\psfrag{c}[c][c][0.8]{$\mathbf{J}_s$}
\psfrag{b}[r][c][0.8]{$\mathbf{E}_{s,1},~\mathbf{H}_{s,1}$}
\psfrag{d}[c][c][0.8]{\shortstack{$\mathbf{E}_{s,2}=0$\\ $\mathbf{H}_{s,2} = 0$}}
\psfrag{e}[c][c][0.8]{$\epsilon_1,~\mu_1$}
\psfrag{f}[c][c][0.8]{$\epsilon_2,~\mu_2$}
\psfrag{g}[c][c][0.8]{$\mathbf{K}_s$}
\psfrag{h}[c][c][0.8]{$\mathbf{J}_s$,~$\mathbf{K}_s$}
\psfrag{j}[c][c][0.8]{$\mathbf{E}_{s,2},~\mathbf{H}_{s,2}$}
\includegraphics[width=0.8\columnwidth]{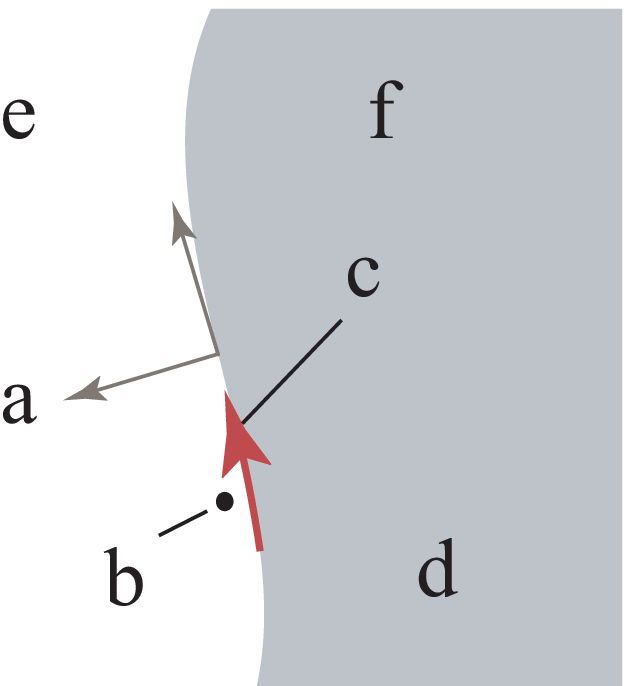}
\caption{}
\end{subfigure}  
\begin{subfigure}[b]{0.23\textwidth}
\centering
\psfrag{a}[r][c][0.8]{$\Jv^s_\text{so},~\Ev^s_\text{so}$}
\psfrag{b}[r][c][0.8]{$\Jv=0$}
\psfrag{c}[l][c][0.8]{$\Jv^s_\text{pec},~\Ev^s_\text{pec}$}
\psfrag{d}[c][c][0.8]{$\E_s,~\H_s$}
\includegraphics[width=\columnwidth]{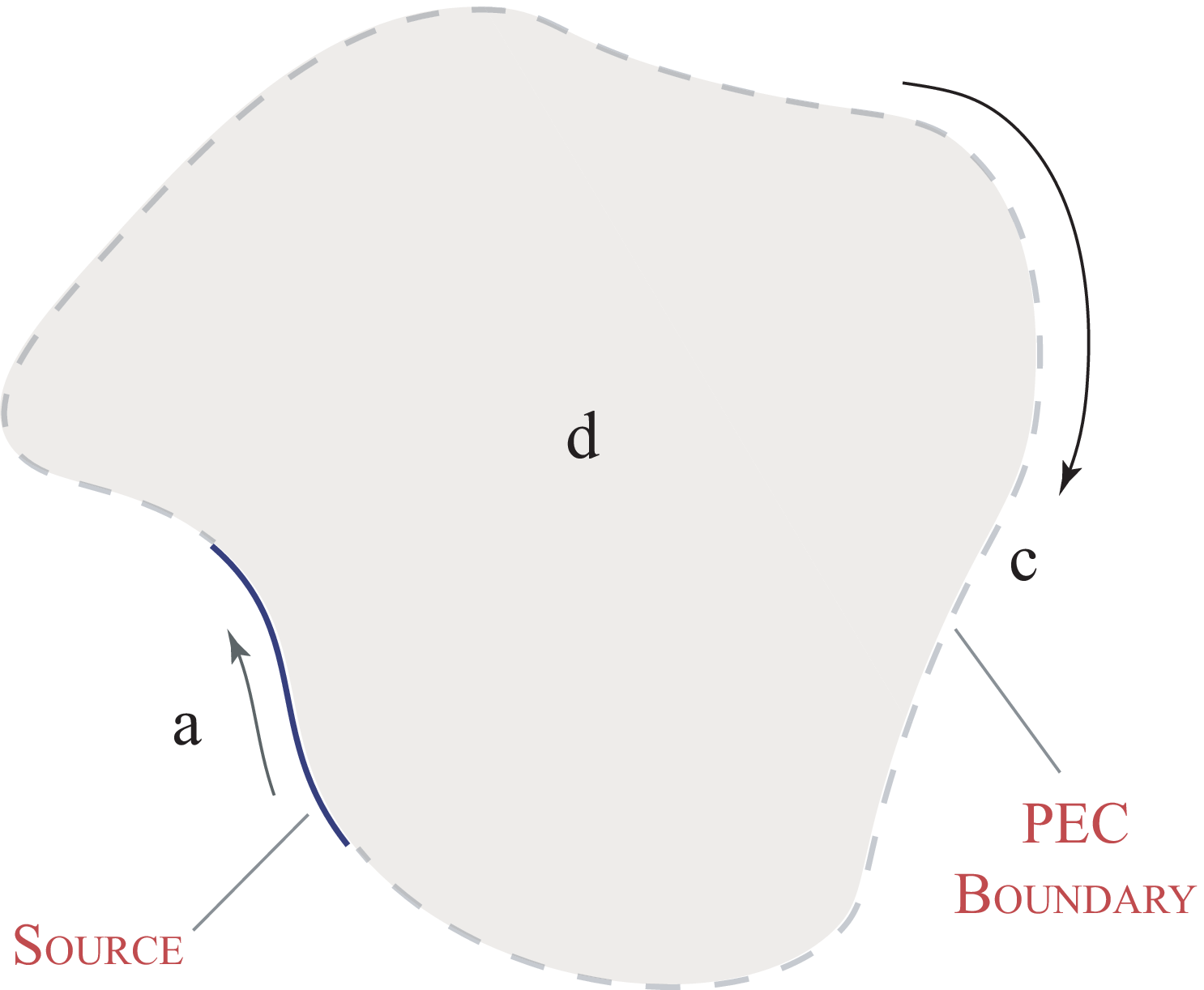}
\caption{}
\end{subfigure}  
\caption{Scattering from a PEC boundary. a) Boundary conditions on a PEC, b) PEC object in the presence of impressed sources.}\label{Fig:PEC_Scat}
\end{figure}

For many EM problems it is not needed to solve simultaneously for the $\E$ and $\H$ fields.  Such a formulation is known as Electric Field Integral Equation (EFIE) which calculates the radiated \emph{electric fields} obtained from the induced surface currents \cite{Method_Moments} \cite{EFIE}. In such cases, the EFIE would enforce Eq.~\eqref{Eq:GenBC} while assuming that $\Kv_s=0$. Both Eq.~\eqref{Eq:GenBC} and Eq. (\ref{Eq:BEMScatteredField}a) are then solved self-consistently for the unknown local current $\Jv^s$ created by known incident fields. For example, simple scattering of an EM wave from PEC boundary, as shown in Fig.~\ref{Fig:PEC_Scat}, will not generate any $\K_s$ terms and only Eq. \ref{Eq:ScatteredEField} needs to be solved in conjunction with the appropriate BCs of the PEC,
 \begin{subequations}\label{Eq:PEC}
\begin{align}
    \N \Ev^s &= 0\\
	\N  \Hv^s &= -\Jv^s.
\end{align}
\end{subequations}

\noindent In this case, we can form a set of equations from geometry consisting of a single source with surface currents $\Jv^s_\text{so}$, surface fields $\Ev^s_\text{so}$ and a single PEC surface with surface currents $\Jv^s_\text{pec}$ and fields $\Ev^s_\text{pec}$. The complete set of surface currents would be $\Jv^s = [ \Jv^s_\text{so} \; \Jv^s_\text{pec}]$ and the surface fields $\Ev^s = [ \Ev^s_\text{pec} \; \Ev^s_\text{so}]$. Using Eq. (\ref{Eq:BEMScatteredField}), (\ref{Eq:Sou}) and (\ref{Eq:PEC}), we can form the field matrix equation:
\begin{align*}
\left[
    \begin{array}{cccc}
    &  & \I & \0 \\
    \multicolumn{2}{c}{\smash{\raisebox{.5\normalbaselineskip}{\large $\left[j\omega\mu\SLm\right]$}}} &  0 & \I \\
    \0 & \0 & \N & \0 \\
    \0 & \0 & \0 & \I \\
    \end{array}
\right]
\left[
    \begin{array}{c}
    \Jv^s_\text{so}\\ \Jv^s_\text{pec}\\\Ev^s_\text{pec}\\\Ev^s_\text{so}
    \end{array}
\right] 
=
\left[
    \begin{array}{c}
    \0 \\ \0 \\ \0 \\ \Ev_0
    \end{array}
\right] 
\end{align*}
or
\begin{align*}
    \Sm \Ym = \Bm,
\end{align*}
\noindent where $\0$ is a Null vector or matrix and $\I$ is an identity matrix. This is a complete set of linear equations that can be solved for $\Jv^s$ and $\Ev^s$, embedded in $\Ym$, from which the fields in the entire region can be calculated using Eq.~(\ref{Eq:SimpleScatteredField}) and the operators given in Eq. (\ref{Eq:BEMOper}).

An alternative formulation, the Magnetic Field Integral Equation (MFIE), calculates the radiated \emph{magnetic fields} from the induced surface currents \cite{Method_Moments}\cite{MFIE}. The MFIE is a Fredholm integral equation of the second kind, compared to the EFIE which is of the first kind, however it is generally limited to closed structures \cite{MFIE}. This formulation would solve the dielectric boundary condition by solving both Eq.~\eqref{Eq:GenBC} and (\ref{Eq:BEMScatteredField}b) for the unknown local current $\J_s$, once again imposing that $\K_s=0$. The two methods can be used using a linear combination to eliminate singularities in the formulation which is known as the Combined Field Integral Equation (CFIE)\cite{chew2009integral}. 

It should be noted that for a general metasurface boundary condition, relating arbitrary fields across it and described in terms of tensorial surface susceptibilities, both surface currents $\J_s$ and $\K_s$ exist on the interface, so that both EFIE/MFIE and CFIE methods cannot be straightforwardly applied to this boundary. Consequently we will develop an appropriate formulation of the metasurface boundary next combining GSTCs and the conventional BEM technique.

\section{Metasurface Integration in BEM}

\subsection{Discretized GSTCs}

In order to simulate the behavior of a general metasurface acting as a boundary between two adjoining regions of space, GSTCs of Eq.~\ref{Eq:FullGSTC} have to be implemented into the BEM equations. Similar to other boundary conditions, the GSTCs relate the electric and magnetic fields on either side of the boundary with each other through the surface susceptibilities. For the sake of simplicity, let us consider a mono-isotropic surface, so that $\overline{\overline{\chi}}_\text{em} = \overline{\overline{\chi}}_\text{me} = 0$ and purely scalar susceptibilities, i.e. $\overline{\overline{\chi}} = \chi$. For such a surface, the metasurface interface condition of Eq.~\eqref{Eq:ExpandedGSTC} for the $i$'th element when discretized are,
\begin{align*}
	&\nh_i \times (\H_{s,1}^i - \H^i_{s,2})
	- \frac{j\omega\epsilon\chi_\text{ee,i}}{2}\left(\E_{s,1}^i+\E_{s,2}^i\right) = 0\\
	&\nh_i \times (\E_{s,1}^i - \E_{s,2}^i) 
	+\frac{j\omega\mu_0\chi_\text{mm,i}}{2}\left(\H_{s,1}^i+\H_{s,2}^i\right) = 0,
\end{align*}
\noindent where the all the fields involved are the \emph{tangential} fields only to the surface, which can further be expressed in matrix form as:
\begin{subequations}\label{Eq:MatGSTC}
\begin{align}
    &\N (\Hv^{1}_\text{m} - \Hv^{2}_\text{m} ) 
    	- \Chibb_e\left(\Ev^{1}_\text{m} +\Ev^{2}_\text{m} \right) = 0 \\
	&\N  (\Ev^{1}_\text{m}  - \Ev^{2}_\text{m} ) 
	+\Chibb_m\left(\Hv^{1}_\text{m} +\Hv^{2}_\text{m} \right) = 0.
\end{align}
\end{subequations}
\noindent where the subscript, $m$, is used to indicate the terms related to metasurface. These equations assume that the fields exit in all three cardinal directions on the surface of the boundary. However, if the surface element's normal vector is in the same direction as one of the spatial directions the GSTCs will only provide two valid equations instead, related to the tangential fields only. In these cases, the relevant equation is removed and another equation is introduced which enforces that the surface current density perpendicular to the surface is zero.

\subsection{Two Region Implementation with Single Metasurface}
\begin{figure}[tbp]
\begin{center}
\psfrag{x}[c][c][0.9]{$x$}
\psfrag{y}[c][c][0.9]{$y$}
\psfrag{a}[c][c][0.8]{$-\infty$}
\psfrag{b}[c][c][0.8]{$d_s$}
\psfrag{c}[c][c][0.8]{$\ell_s$}
\psfrag{d}[r][c][0.8]{$\Ev^1_\text{m},~\Hv^1_\text{m}$}
\psfrag{e}[l][c][0.8]{$\Ev^2_\text{m},~\Hv^2_\text{m}$}
\psfrag{f}[l][c][0.8]{\shortstack{$\Jv_\text{so},~\Kv_\text{so}$\\$\Ev_\text{so},~\Hv_\text{so}$}}
\psfrag{g}[r][c][0.8]{$\Jv_\text{m},~\Kv_\text{m}$}
\psfrag{h}[c][c][0.8]{Eq.~\eqref{Eq:IE_2Reg}}
\psfrag{j}[c][c][0.8]{Eq.~\eqref{Eq:MatGSTC}}
\includegraphics[width=\columnwidth]{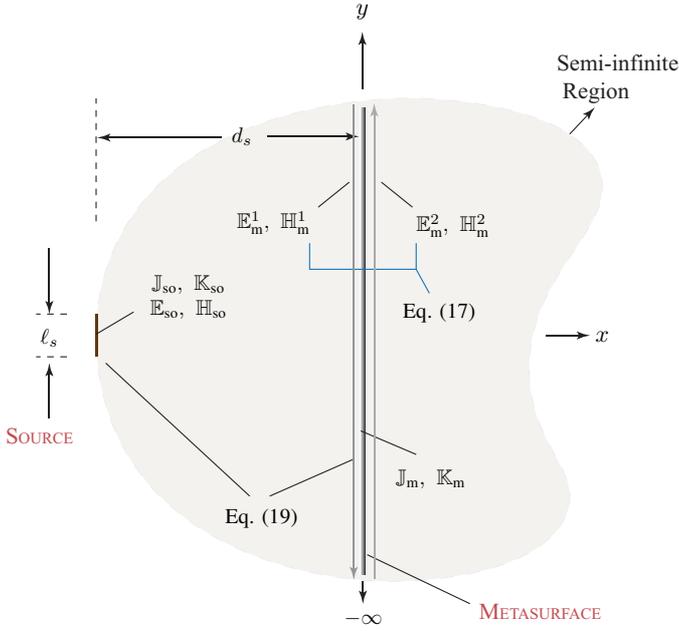}
\caption{Problem of two semi-infinite regions separated by a single metasurface described in terms of its scalar surface susceptibilities, where a source is specified in region 1 on the left.}\label{Fig:SimSetup}
\end{center}
\end{figure}

\begin{figure*}[t]
\begin{align*}
\begin{array}{c}
\hline
\\
\\
\smash{\raisebox{.4\normalbaselineskip}{ Eq. \ref{Eq:IE_2Reg}a and b}}\\
\\
\hline
\\
\smash{\raisebox{.4\normalbaselineskip}{ Eq. \ref{Eq:IE_2Reg}c and d}}\\
\hline
\\
\smash{\raisebox{.4\normalbaselineskip}{ Eq. \ref{Eq:MatGSTC}}}\\
\hline
\\
\smash{\raisebox{.4\normalbaselineskip}{ Eq. \ref{Eq:Sou}}}
\\
\hline
\end{array}
\left. 
\begin{array}{c}
    \\ \\ \\ \\ \\ \\ \\ \\ \\ \\
\end{array} 
    \right\}
    \Longrightarrow 
\left[
    \begin{array}{cccccccccc}
     &  &  &  & \I & \0 & \0 & \0 & \0 & \0 \\
    \multicolumn{2}{c}{\smash{\raisebox{.5\normalbaselineskip}{\large $\left[j\omega\mu\SLmOne\right]$}}} &
    \multicolumn{2}{c}{\smash{\raisebox{.5\normalbaselineskip}{\large $\left[\SRmOne\right]$}}}  & \0 & \0 & \0 & \0 & \I & \0 \\
     &  &  &  & \0 & \I & \0 & \0 & \0 & \0 \\
    \multicolumn{2}{c}{\smash{\raisebox{.5\normalbaselineskip}{\large $\left[-\SRmOne\right]$}}} &
    \multicolumn{2}{c}{\smash{\raisebox{.5\normalbaselineskip}{\large $\left[j\omega\epsilon\SLmOne\right]$}}}  & \0 & \0 & \0 & \0 & \0 & \I \\
    \0 & j\omega\mu\SLmTwo & \0 & \SRmTwo & \0 & \0 & \I & \0 & \0 & \0 \\
    \0 & -\SRmTwo & \0 & j\omega\epsilon\SLmTwo & \0 & \0 & \0 & \I & \0 & \0 \\
    \0 & \0 & \0 & \0 & \N & \Chibb_\text{m} & -\N & \Chibb_\text{m} & \0 & \0 \\
    \0 & \0 & \0 & \0 & -\Chibb_\text{e} & \N & -\Chibb_\text{e} & -\N & \0 & \0 \\
    \0 & \0 & \0 & \0 & \0 & \0 & \0 & \0 & \I & \0 \\
    \0 & \0 & \0 & \0 & \0 & \0 & \0 & \0 & \0 & \I \\
    \end{array}
\right]\nonumber
\left[
    \begin{array}{c}
    \Jv_\text{so}\\ \Jv_\text{m}\\\Kv_\text{so}\\ \Kv_\text{m}\\ \Ev^1_\text{m}\\\Hv^1_{m}\\\Ev^2_\text{m}\\\Hv^2_\text{m}\\\Ev_\text{so}\\\Hv_\text{so}
    \end{array}
\right] 
= 
\left[
    \begin{array}{c}
    \0\\ \0\\ \0 \\ \0 \\ \0 \\ \0 \\ \0 \\ \0  \\ \Ev_0 \\ \Hv_0
    \end{array}
\right] \nonumber
\end{align*}
\begin{equation}
\Sm \Ym = \Bm \label{Eq:TwoRegScat}
\end{equation}
\end{figure*}

To illustrate the introduction of the metasurface into a BEM method, we will present for simplicity the mathematical formulation for two regions only. The first with a single source and the two regions simply connected by a metasurface as shown in Fig. \ref{Fig:SimSetup}. In the first region, the surface currents consist of $\Jv^s_1 = [\Jv_\text{so},\; \Jv_\text{m}]$ and $\Kv^s_1 = [\Kv_\text{so},\;\Kv_\text{m}]$ and for the second region where only the metasurface is present we have $\Jv_s^2 = \Jv_\text{m}$ and $\Kv_s^2 = \Kv_\text{m}$. For each of the regions, the operators $\SLm$ and $\SRm$ are formed noting that $\Jv_\text{m}$ and $\Kv_\text{m}$ are present for both, leading to
\begin{subequations}\label{Eq:IE_2Reg}
\begin{align}
	\Ev^s_1 = -j\omega\mu \SLmOne \Jv^s_1 - \SRmOne \Kv^s_1 \\ 	
	\Hv^s_1 = -j\omega\epsilon \SLmOne \Kv^s_1 + \SRmOne \Jv^s_1 \\ 	
	\Ev^s_2 = -j\omega\mu \SLmTwo \Jv^s_2 - \SRmTwo \Kv^s_2\\
	\Hv^s_2 = -j\omega\epsilon \SLmTwo \Kv^s_2 + \SRmTwo \Jv^s_2.
\end{align}
\end{subequations}
\noindent Note that the ``open" radiating surface is not included in the discretized surface model as \tjs{it is at infinity}. The GSTC expressed in Eq.~(\ref{Eq:MatGSTC}) are then used to related the field $\Ev^1_\text{m}$, $\Ev^2_\text{m}$, $\Hv^1_\text{m}$ and $\Hv^2_\text{m}$, and the source equation (\ref{Eq:Sou}) to define the source fields $\Ev_\text{so} = \Ev_0$ and $\Hv_\text{so} = \Hv_0$. Placing all these equations in a matrix formulation produces Eq. (\ref{Eq:TwoRegScat}).

\noindent Which can now finally be solved for all the unknown surface currents and fields and Eq.~\eqref{Eq:SimpleScatteredField} can be subsequently used to calculate fields anywhere within the two regions. 

\section{Numerical Demonstration}

\subsection{Simulation Setup}

Fig.~\ref{Fig:SimSetup} shows the numerical setup consisting of a metasurface of length $\ell$ located at $x=0$. In order to simplify the simulation, a 2D problem is considered where the field varies only in the $x-y$ plane. \tjsrm{A source surface is placed parallel to the metasurface at $x=-d_{s}$ with a length of $\ell_s$ and is modeled using an implementation of a dielectric boundary condition.}
\tjs{An input source consisting of both $\E_s$ and $\H_s$ fields is applied at $x=-d_{s}$ using a source surface of length $\ell$. This source is configured to create a TE field with a diffracting Gaussian-like profile with the waist at $d_s$ and width of $\sigma_y$.}

It should be noted that the separation between the source and the metasurface, $d_s$ does not affect the memory requirements of the simulation since the surfaces are linked together using the scattered field equations Eq.~\eqref{Eq:BEMScatteredField}. In addition, the source and surface discretization can also be different. This implies that the computation region can be arbitrarily large. 

The frequency dependence of metasurface susceptibilities are assumed to follow a Lorentzian distribution, given by
\begin{subequations}\label{Eq:DualLZ}
\begin{equation}
	{\chi}_\text{ee}(\omega) =\frac{\omega_{ep}^2}{(\omega_{e0}^2 - \omega^2) + j\alpha_{e}\omega}  
\end{equation}
\begin{equation}
	{\chi}_\text{mm}(\omega) = \frac{\omega_{mp}^2}{(\omega_{m0}^2 - \omega^2) + j\alpha_{m}\omega}, 
\end{equation}
\end{subequations}
\noindent where $\omega_p$, $\omega_0$ and $\alpha$ are the plasma frequency, resonant frequency and the loss-factor of the oscillator, respectively, and subscripts $e$ and $m$ denote electric and magnetic quantities. For the chosen operation frequency of 60~GHz, the metasurface size is fixed to $\ell=0.1~$m and the source surface's length is set to $\ell_s=0.08~$m with a separation of $d_{s}=0.05~$m. \tjs{The metasurface length was chosen to be sufficiently large that the source field at the surface was contained within it.}

In our numerical simulation, the discretization of the metasurface and the source was set to $n_\lambda=40$ divisions per wavelength.  The fields in a rectangular region surrounding the metasurface with dimensions $x = [-d_s,~d_s]$ and $y = [-\ell/2,~\ell/2]$ and discretization $\Delta x = \lambda/10$ and $\Delta y = \lambda/n_\lambda$ were calculated using Eq.~\eqref{Eq:BEMScatteredField} after the initial simulation. 

\tjs{One advantage of the BEM method is the ability to create visualizations of the field distributions of either the total field (excitation plus scattered fields) or to visualize these fields independently. For example in the first region, one can calculate the incident field by simply applying Eq. \ref{Eq:BEMScatteredField} to the currents present in the source surface. Conversely, for either region, Eq. \ref{Eq:BEMScatteredField} can be applied to the currents in the metasurface and the scattered fields (reflected and transmitted) can be determined. Of course, if in the first region both source and metasurface currents are used, then total fields will be calculated.}

\subsection{Simulation Results}

\begin{figure}
\centering
\psfrag{a}[c][c][0.7]{$y$~(m)}
\psfrag{b}[c][c][0.7]{$x$~(m)}
\psfrag{e}[c][c][0.8]{$20\log|E_\text{total}(x,y)|$~dB}
\psfrag{d}[c][c][0.8]{$20\log|E_\text{scat}(x,y)|$~dB}
\psfrag{c}[c][c][0.8]{$20\log|E_\text{inc}(x,y)|$~dB}
\includegraphics[width=\columnwidth]{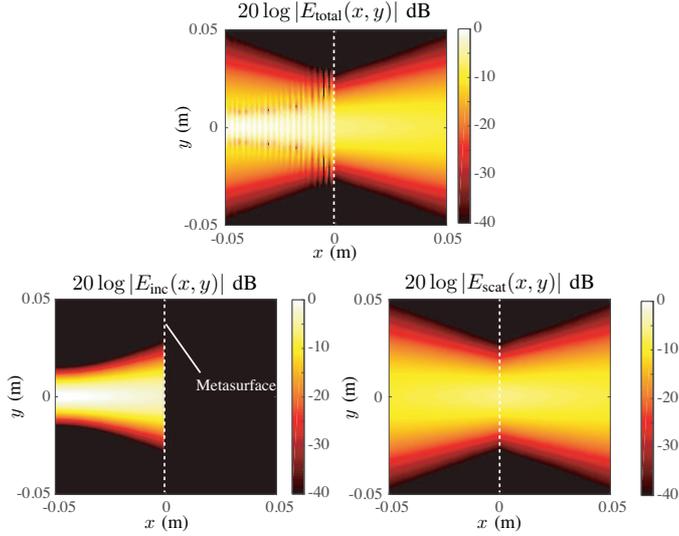}
\caption{Scattered field solution of the two semi-infinite regions connected by a single metasurface. a) 2D E-filed distribution of the incident, scattered and total E-fields in both regions. Metasurface surface susceptibilities were chosen to provide a strong interaction at 60~GHz following the Lorentzian model: $\omega_{p,e}=\omega_{p,m} = 9\times10^9~$rad/s,  $\omega_{r0,e} = 2\pi\;57\times10^9~$rad/s, $\omega_{r0,m}=2\pi\;37\times10^9~$rad/s, and $\alpha_e=\alpha_m = 2\pi\times10^9$.}\label{Fig:Results1}  
\end{figure}

\begin{figure}
\begin{subfigure}[b]{0.5\textwidth}
\centering
\psfrag{a}[c][c][0.7]{$y$~(m)}
\psfrag{b}[c][c][0.7]{$|E_r(y, x = 0_-)|$}
\psfrag{c}[c][c][0.7]{$|E_t(y, x = 0_+)|$}
\psfrag{d}[l][c][0.6]{$n_\lambda = 5$}
\psfrag{e}[l][c][0.6]{$n_\lambda = 10$}
\psfrag{f}[l][c][0.6]{$n_\lambda = 100$}
\includegraphics[width=\columnwidth]{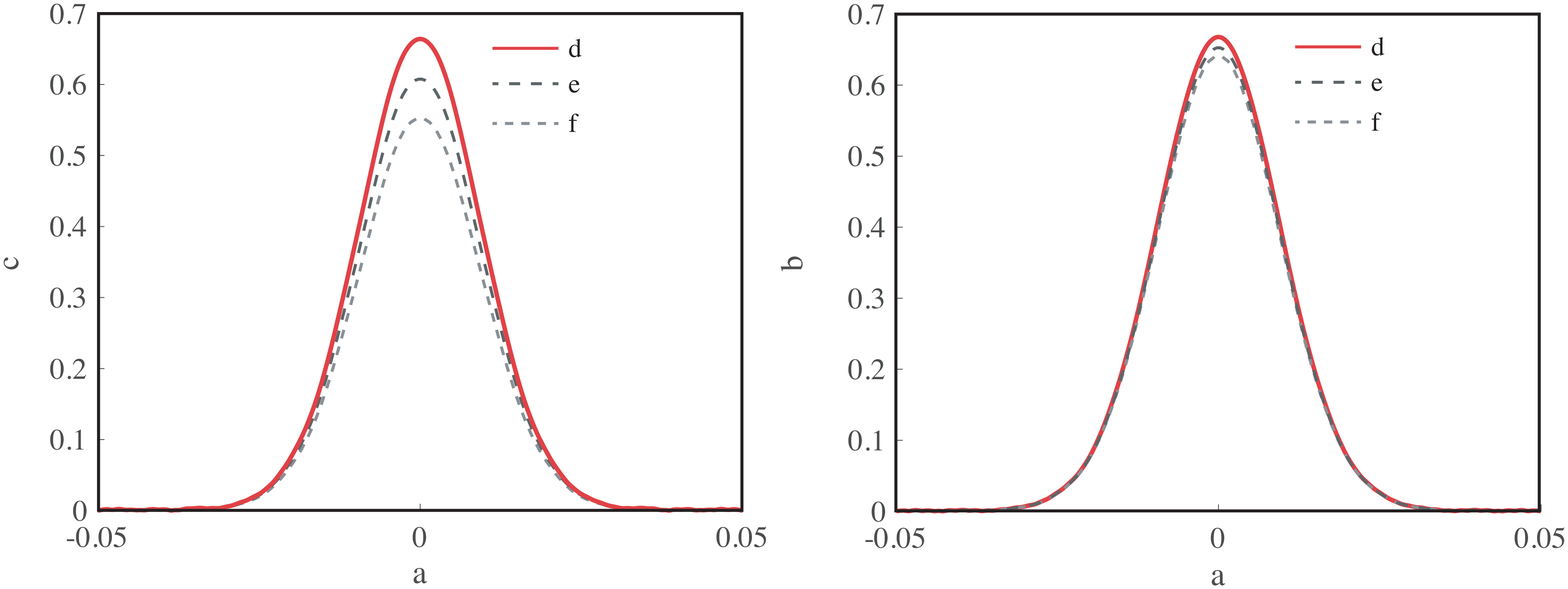}

\end{subfigure}
\begin{subfigure}[b]{0.5\textwidth}
\centering
\psfrag{a}[c][c][0.7]{divisions per wavelength, $n_\lambda$}
\psfrag{b}[c][c][0.7]{$|E_t(y=0, x = 0_+)|$}
\psfrag{g}[c][c][0.7]{$|E_r(y=0, x = 0_+)|$}
\psfrag{c}[c][c][0.6]{analytical}
\psfrag{d}[c][c][0.6]{numerical}
\psfrag{e}[c][c][0.8]{Transmission Region}
\psfrag{f}[c][c][0.8]{Reflection Region}
\includegraphics[width=\columnwidth]{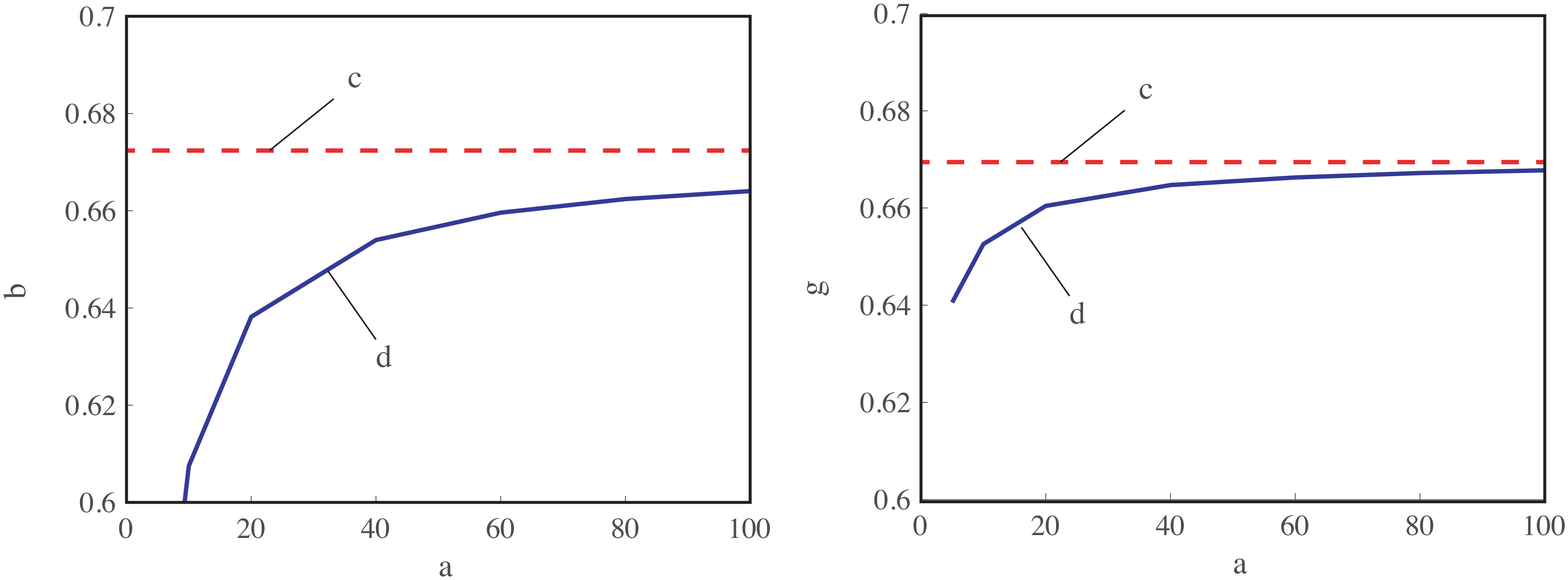}
\caption{}
\end{subfigure}
\caption{Convergence of the transmitted and reflection fields in Fig. \ref{Fig:Results1} as a function of the mesh density.}\label{Fig:Results2}
\end{figure}

To demonstrate the method, two cases of metasurfaces will be considered: 1) A uniform metasurface with $\chi_\text{ee} \ne \chi_\text{mm}$, and 2) a non-uniform metasurface with spatially varying $\chi_\text{ee}(y)= \chi_\text{mm}(y)$.

The uniform metasurface is chosen as the first example, since its scattered fields can be readily determined analytically. For a uniform metasurface excited with an arbitrary shaped input beam $\E_0(y)$, at a fixed angular frequency, $\omega$, the transmitted and reflected fields in the spatial frequency $k_y$, are given by, \cite{Stewart_Uniform_SemiAna}
\begin{subequations}
\begin{align}
\tilde{\mathbf{E}}_t(k_y)   &=  {\left[ \frac{4 + k_0^2{\chi}_\text{ee}{\chi}_\text{mm}}{(2+ jk_x{\chi}_\text{mm})(2 +  j(k_0^2/k_x){\chi}_\text{ee})}\right]}\tilde{\mathbf{E}}_0(k_y)\\
\tilde{\mathbf{E}}_r(k_y)   &=  {\left[ \frac{2j(k_x{\chi}_\text{mm} - (k_0^2/k_x){\chi}_\text{ee})}{(2+ jk_x{\chi}_\text{mm})(2 +  j(k_0^2/k_x){\chi}_\text{ee})}\right]}\tilde{\mathbf{E}}_0(k_y)
\end{align}
\end{subequations}
\noindent where $\tilde{\mathbf{E}}(k_y)$ represents the spatial Fourier transform of $\mathbf{E}(y)$, \sg{$\mathbf{E}_t$ and $\mathbf{E}_t$ are the scattered fields in transmission and reflection}. This physically represents the transmitted and reflected field response of the metasurface for a specific $k_y$, which corresponds to a specific plane-wave excitation (in the propagation regime). The spatial scattered fields $\tilde{\mathbf{E}}_t(y)$ and $\tilde{\mathbf{E}}_r(y)$ are obtained using inverse Fourier transforms of the above fields: $\mathbf{E}_t(x = 0+,y) = \mathcal{F}_y^{-1}\left\{\tilde{\mathbf{E}}_t(k_y)\right\}$ and $\mathbf{E}_r(x = 0_-,y) = \mathcal{F}_y^{-1}\left\{\tilde{\mathbf{E}}_r(k_y)\right\}$. This method can be used to validate the numerical results of the proposed method.

Fig.~\ref{Fig:Results1} shows the total scattered fields and the scattered fields just before and after the metasurface. The convergence plots in Fig.~\ref{Fig:Results2} show the effect of meshing density on the computed fields, where both the transmitted and reflected fields are clearly seen to be converging to analytical values, which indicates that with higher discretization they should approach even closer to the expected results. This provides a good validation of the method. 

\begin{figure}
\centering
\psfrag{g}[c][c][0.8]{$20\log|E_\text{total}(x,y)|$~dB}
\psfrag{a}[c][c][0.8]{$y~$(m)}
\psfrag{b}[c][c][0.8]{$x~$(m)}
\psfrag{d}[c][c][0.8]{$k_y~\text{m}^{-1}$}
\psfrag{c}[c][c][0.8]{$\mathcal{F}\{E_t(y)\}$ (normalized)}
\psfrag{f}[c][c][0.7]{FDFD~\cite{CalozFDTD}}
\psfrag{e}[c][c][0.7]{BEM-GSTC}
\includegraphics[width=1\columnwidth]{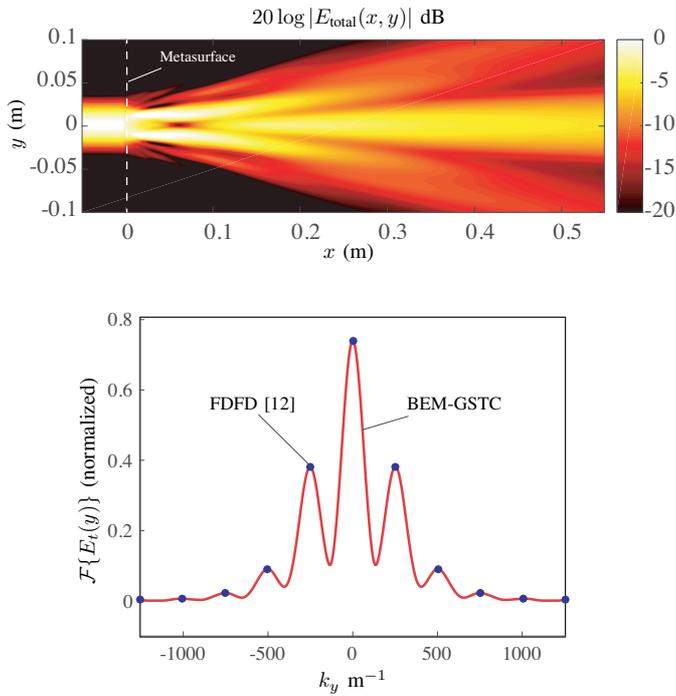}\label{Fig:Grating2D}
\caption{2D Scattered fields from a metasurface diffraction grating and the comparison between the spatial Fourier transform of the transmitted fields with an FDFD based method of ~\cite{CalozFDTD}. Metasurface surface susceptibilities are the same as in Fig.~\ref{Fig:Results1}, except $\chi_\text{mm}(y) = \chi_\text{ee}(y)$ and $\sigma_y=4/\beta_m$. Mesh density $n_\lambda = 100$.}\label{Fig:Grating}
\end{figure}

Next, a non-uniform metasurface is considered whose electric and magnetic surface susceptibilities are assumed equal but modulated in space, emulating a diffraction grating. For a physical metasurface, this can be achieved by periodically modulating the resonant frequency $\omega_0$ of the Lorentzian function of Eq.~\eqref{Eq:DualLZ}, given by
\begin{equation}
	\omega_{0}(y) = \omega_{r0,q} \left\{1+\Delta_m\cos\left(\beta_m y\right)\right\}
\end{equation}
\noindent where $\Delta_m$ controls the intensity of the modulation and $\beta_m$ is the spatial frequency of modulation. To produce strong diffraction orders, $\Delta_m=0.1$ and $\beta_m=k/5$ were used with a Gaussian beam of width $\sigma_y=4/\beta_m$. Fig.~\ref{Fig:Grating}a shows the calculated total power in the two regions where the normally-incident Gaussian beam is split into several diverging higher-order diffraction order beams. To better visualize these diffraction orders, the transmitted field of the metasurface is captured and a spatial Fast Fourier Transform (FFT) is applied. Fig.~\ref{Fig:Grating}b shows the strengths of various diffraction orders as a function of the mesh density, $n_\lambda$. Resulting diffractions orders are first seen to be equally spaced with $k_y=\beta_m$ as expected, and their strengths are gradually converged to a constant value beyond which higher meshing has no impact on the results. To validate these results, the strengths of the generated harmonics are compared with a Yee-cell based Finite Difference Frequency Domain (FDFD) method of \cite{CalozFDTD}, using the same metasurface parameters with a discretization of $n_\lambda=40$. Fig.~\ref{Fig:Grating}b also shows this comparison, where an excellent agreement is observed between the FDFD and the proposed method, so that the validity of the method is clearly established.

\section{Conclusion}

\tjs{A methodology for determining the scattered EM fields present for interconnected regions with metasurface boundaries has been proposed. The method is based on a BEM formulation of the frequency domain version of Maxwell's equations -- which expresses the fields present in a region due to surface currents on the boundaries. Multiple regions are coupled together by shared surface currents which can be solved for in a self-consistent manner. The general metasurface boundaries represented as surface susceptibilities were integrated in the BEM method next using GSTCs.  }


\tjs{To illustrate the method, two examples were presented for EM scattering of a Gaussian beam. Firstly a uniform metasurface, and secondly a spatially modulated metasurface. For the uniform surface the BEM results were compared to a semi-analytical method and it was shown that as the boundary segment length was decreased, the BEM results converged to the semi-analytical results. The second example was for a spatially modulated metasurface (essentially a grating) and the BEM results were compared to a Yee-cell frequency domain method. The two methods were found to predict essentially the same diffraction components confirming the BEM accuracy. It can be noted that the BEM method as formulated used a very simple uniform element function and more sophisticated methods would allow for larger elements to be used. However, for this paper the purpose is to simply show the functionality of the method. }

\tjs{Although, this paper shows a simple application of the BEM method with an incorporated metasurface, it establishes the applicability of the method for larger scale scattering problems. Moreover, the GSTC model of metasurface boundaries may also be seen as pure numerical tool to connect multiple regions by mimicking and transcending the functionalities of conventional boundaries. Situations where the metasurface is a part of a larger problem with multiple scattering objects (both electrically large and small), and scattering response is the prime objective (similar to Radar Cross-Section, RCS), the proposed approach may prove to be an invaluable tool. Such problems have been a traditional domain for BEM modeling and with the appropriate incorporation of advanced BEM methods and computational techniques, the proposed methodology is likely to be increasingly useful.}





%

\end{document}